\documentstyle[preprint,tighten,aps]{revtex} 

\begin{document}

\draft \title{ Mapping vesicle shapes into the phase diagram:\\
 A comparison of experiment and theory.}

\author{ H.-G. D\"{o}bereiner$^{\ast }$,
              E. Evans$^{\dagger }$,
              M. Kraus$^{^\ast }$,
              U. Seifert$^\ast $, and
              M. Wortis$^{\ddagger}$}

\address{
$^\ast$Max-Planck-Institut f\"{u}r Kolloid-
und Grenzfl\"{a}chenforschung,\\
Kantstrasse 55, 14513 Teltow-Seehof, Germany\\
$^\dagger$Department of Physics, University of British Columbia,\\
 6224 Agriculture Road, Vancouver, British Columbia, Canada V6T 2A6\\
 $^\ddagger$Physics Department, Simon Fraser University,\\
Burnaby, British Columbia, Canada V5A 1S6\\
}

\maketitle

\begin{abstract}

Phase-contrast microscopy is used to monitor the
shapes of micron-scale fluid-phase phospholipid-bilayer
vesicles in aqueous solution.
At fixed temperature, each vesicle undergoes thermal shape fluctuations.
We are able
experimentally to characterize the
thermal shape ensemble by digitizing the vesicle
outline in real time and storing the time-sequence of images.
Analysis of
this ensemble using the area-difference-elasticity (ADE) model
of vesicle shapes allows us to associate (map) each time-sequence
to a point in the zero-temperature (shape) phase diagram.
Changing the laboratory temperature modifies the control
parameters (area, volume, etc.) of each vesicle, so it
sweeps out a trajectory across the theoretical phase diagram.
It is a nontrivial test of the ADE model to check that
these trajectories remain confined to regions of the phase
diagram where the corresponding shapes are locally stable.  In
particular, we study the thermal trajectories of three prolate vesicles
which, upon heating, experienced a mechanical instability
leading to budding.  We verify that the position
of the observed instability and the geometry of the budded
shape are in reasonable accord with the theoretical
predictions.  The inability of previous experiments to detect the
``hidden'' control parameters (relaxed area difference and
spontaneous curvature) make this the first direct quantitative
confrontation between vesicle-shape theory and experiment.

\vspace{1cm}\mbox{} \pacs{\noindent submitted to PRE}

\end{abstract}

\def\lsim{\ \raise 1pt\hbox{$<$}\lower 3pt\hbox{\llap{$\sim$}}\ }
\def\gsim{\ \raise 1pt\hbox{$>$}\lower 3pt\hbox{\llap{$\sim$}}\ }

\def\cel{$^\circ$\ C}



\section{Introduction}
\label{intro} Micron-scale fluid-phase
lipid-bilayer vesicles have been observed in recent years under controlled
laboratory conditions
\cite{Lipo91,Sack86a,Sack86b,Evan90,Duwe90,Bern90,Kaes91,Farg92,Doeb93,Doeb95,Wint95}
to
exhibit many amusing and diverse shapes.  At the same time, there is now a
one-parameter theory
of vesicle shapes, the
so-called area-difference-elasticity (ADE) model
\cite{Miao94,Svet82,Svet83,Svet89},
which
appears to be qualitatively consistent with available
experimental observations.  It would be nice, however, to have a vesicle
with an
accurately measured shape and known parameters, to plug these parameters
into
the theory, to predict a shape, and to compare with the measured one.  Up
to
this time, this has not been possible, and, indeed, there have been few
(if any)
direct quantitative confrontations between theory and experiment.

The reasons
for this unsatisfactory state of affairs have their origins in both theory
and
experiment.   On the theoretical side, the principal models which have been
proposed to describe vesicle shapes
\cite{Miao94,Seif91,Four94,Seif96}
all have in common the same catalogue
of stationary-energy shapes.  Thus, simple observation of a vesicle whose
shape can be found in the catalogue, while evidence for the general
validity of
bending-energy models, does not distinguish one variant from another.
In
order to test the model, it is necessary to study more indirect and/or
delicate
issues such as stability (absolute and relative) or shape-change
systematics
under variation of control parameters.  This has not often been done
\cite{comment1}
for fundamental experimental reasons:    First, there are two
important vesicle parameters which can be modified systematically in the
lab
but are not subject to direct measurement.  One of these is the
spontaneous curvature, $C_0$, which measures the preferred radius of
curvature
of the relaxed bilayer, based on the different lipid composition of the two
constituent monolayer leaves and/or the different aqueous environments
inside
and outside the vesicle.  This parameter is presumably the same for all
vesicles
in a single homogeneous suspension.   The other is the relaxed area
difference,
$\Delta A_0$, between the two leaves, based on the different number of
lipid
molecules which they contain \cite{Mui95} and the long relaxation time for lipid
exchange
between them \cite{Korn71,Homa88}.  This parameter will in general
vary
from one vesicle to another in the same suspension, based on the (unknown)
manner in which vesicle closure occured during preparation and on any
interleaf ``flip-flop''
or intercalation events which may have
occured subsequently.  In addition, the vesicle shape,
itself, is a significantly ambiguous quantity.  At laboratory
temperatures, all nonspherical
vesicles undergo significant and unavoidable thermal fluctuations.  Thus,
at any
nonzero temperature $T$, experiment must characterize a thermal shape
ensemble.
A single ``snapshot,'' such as has often
been exhibited in the previous literature,
cannot do this.  Finally, experiments have not in practice probed the full,
three-dimensional vesicle shape but at best a two-dimensional cut through
it at
the focal plane of the observing apparatus.

It is the aim of this paper to show how to deal with all
these problems in a serious manner for the first time.  Using video
phase contrast microscopy,
we recorded for each vesicle and at each temperature long time-sequences of
two-dimensional vesicle contours.  We parameterized these images in terms
of a set
of shape amplitudes.  We used the shape-amplitude time-sequences to
construct a
thermal ensemble, from which we extracted a set of thermal expectation
values.
Using this data, we show below how to associate each vesicle with a
particular
point in the phase diagram.  In principle, information is left over after
the
mapping, so that a nontrivial confrontation between theory and experiment
is
possible.  In practice, available resolution limits what we can do;
nevertheless, nontrivial checks are possible.

Overall, the results are
encouraging.  The vesicles that we have located in (``mapped
into'')
the theoretical phase diagram do, generally, end up in regions where they
are predicted to be
locally stable and to have low energies.  Furthermore,
observed
thermal trajectories exhibit shape instability close to (if not always
exactly
at) positions predicted by the theory.  Finally, after the instability
the shape is
in
reasonable agreement with theoretical expectations.

Gravitational effects play an important role
experimentally.  In order to record long time-sequences of vesicle
shapes, it is convenient to adjust the density of the exterior
solution so that the vesicles have a small negative buoyancy and
collect at the floor of the experimental cell, where they
remain within the focal plane of the microscope for long
periods.  In addition, gravity orients the long axis of
vesicles of prolate shape so that it stays in or near the
focal plane.  These are practical issues. On the conceptual
side, whenever it has non-neutral buoyancy, a vesicle is
subject to gravitational shape deformations.  The importance of
these deformations has only recently been recognized \cite{Krau95}
and was not considered in the analysis of earlier experiments.
In this paper, we first perform the full analysis without including
gravity.  Then, we devote a separate section to the
consideration of gravitational corrections.
The upshot is that gravitational effects can
be significant; however, in the region of the phase diagram
upon which we focus attention, there is no change
in the qualitatively good agreement between theory and experiment.

The layout of the paper
is as follows:  Section~\ref{background} introduces the theoretical
background necessary to
analyse the experiments.  Section~\ref{techniques} describes the
experimental procedures.
Section~\ref{shapecontours} explains how the analysis of the
experimental shape contours was carried out.
Section~\ref{results} sets forth our results using a pure ADE
mapping and ignoring gravitational effects.  Finally, in
Section~\ref{gravity}, we explore the effects of gravity.
 Section~\ref{conclusion}
provides a final assessment
and
summary.

\section{Background}
\label{background}

\subsection{The area-difference-elasticity model}
\label{ADEmodel}

In order to have a language for discussing
the experiments, it will be useful to present here a summary of some
principal
features of the ADE model.  Additional material is available elsewhere
\cite{Miao94,Svet82,Svet83,Svet89,Seif96,Miaothesis,HGDthesis}.
At mesoscopic length scales, larger than molecular sizes but
smaller than the persistence length, the shape $S$ of a fluid-bilayer
vesicle
is controlled by an energy functional $W[S]$ consisting of two parts.  The
first, due to Helfrich \cite{Helf73}, measures the overall bending energy and
is
scaled by the bending modulus $\kappa$.  The second requires a brief
explanation:  Assuming fixed bilayer separation, $D$, the actual area
difference
between the two leaves of the bilayer is
\begin{equation}
\Delta A[S]=2D\oint dA H({\bf r}),
\end{equation}
where $H({\bf r})$ is the local mean curvature at the
point ${\bf r}$ of the vesicle surface and the integral runs over the
(closed)
vesicle surface.  On the other hand, the preferred or relaxed area
difference,
\begin{equation}
\Delta A_{0}=(N_{out}-N_{in}) a_{0}(T) ,
\end{equation}
of the two leaves is determined by the difference
$(N_{out}-N_{in})$ between the number of lipid molecules in the outer and
inner
leaves.  The relaxed area $a_{0}(T)$ per lipid molecule is a material
parameter
but can depend, of course, on the temperature  $T$.  Once the vesicle has
closed, $\Delta A_{0}$ can only change due to lipid flip-flop between the
two
leaves and/or lipid interchange with the aqueous environment of the
vesicle,
processes which are believed to be slow on the timescale of the mechanical
shape changes we shall be discussing \cite{Korn71,Homa88}.
The second contribution to $W[S]$
measures the elastic energy necessary to force $\Delta A[S]$ to differ
from $\Delta A_{0}$, when the vesicle assumes the shape $S$.  Because the
vesicle is fluid, this (local) elastic strain is distributed uniformly
over the
vesicle surface and appears as an apparently nonlocal term controlled by a
so-called nonlocal bending modulus $\bar{\kappa}$.  The moduli $\kappa$ and
$\bar{\kappa}$ are both of order $KD^2$, where $K$ is the area stretching
modulus of the bilayer \cite{Miao94}, so the ratio,
\begin{equation}
\alpha \equiv \bar{\kappa}/\kappa ,
\end{equation}
is generically of order unity.  The
material parameters $\kappa$ and $\bar{\kappa}$ can be measured directly.
For
SOPC, it is believed that $\kappa \sim 0.90\pm 0.06\times
10^{-19} {\rm J}$ \cite{Evan87}.
It has been estimated that $\alpha \sim 1.4$ \cite{Miao94}.
(A somewhat higher value, $\kappa \sim 1.20\pm 0.17\times
10^{-19} {\rm J}$ and a comparable but quite
uncertain value of $\alpha$ have been recently observed in
tether-pulling experiments \cite{Waug92}.)
The energy scale $\kappa$ is much smaller than the energies necessary to change
significantly the area $A$ and volume $V$ of the vesicle
\cite{comment3},
so these quantities may be regarded as fixed in comparing
the energies $W[S]$ of different shapes.

Combining the two terms described in
the previous paragraph
(and dropping an irrelevant, shape-independent term)
leads to,
\begin{equation}
\label{ADE}
W[S]=\kappa \left[ G[S] + \frac{\alpha}{2}(m[S]-\bar m_0)^{2}\right] ,
\end{equation}
where
\begin{equation}
G[S] = \frac{1}{2}\oint dA(2H)^2,
\end{equation}
which is the starting point of our theoretical discussion of
shapes.  In writing Eq. (\ref{ADE}), we have chosen to rescale all lengths
in
terms of an ``area length'' $R_A$ defined by $A \equiv 4\pi R_{A}^{2}$.
Thus,
the area difference appears in the reduced form,
\begin{equation}
m[S]=\Delta A/2DR_{A} ,
\end{equation}
and the relaxed area difference combines with the
spontaneous curvature $C_0$ into a single effective reduced area
difference,
\begin{equation}
\label{m0bar}
\bar m_0=m_{0}+2c_{0}/\alpha ,
\end{equation}
where $m_{0}=\Delta A_{0}/2DR_{A}$ and
$c_{0} \equiv C_{0}R_{A}$ is the reduced value of the
spontaneous curvature.  Because $C_{0}$ and $\Delta A_{0}$ appear only in
the
combination $\bar m_0$, it is impossible in  principle to detect either
one
separately by a single shape measurement.  Note that the
bracketed terms
in Eq.~(\ref{ADE}) are all dimensionless ratios of lengths, invariant
under a
scale change of the shape $S$, provided that at the same time $C_0$ is
changed
to keep $c_{0}$ fixed.
In this sense, $W[S]$
depends on the shape of $S$ but not its overall size.  An appropriate
scale-independent volume measure is the reduced volume $v \equiv 3V/4\pi
R_{A}^{3}$, which lies in the interval $[0,1]$.

To be a mechanically viable
shape for a vesicle with given $A$ and $V$, $S$ must make the
energy (\ref{ADE})
stationary at the corresponding values of $v$ and $\bar m_0$
\cite{comment4}, i.e., it must satisfy
\begin{equation}
\label{variation}
\delta W=0=\kappa \Bigl( \delta  G[S] - \alpha (\bar m_0-m[S]) \delta m[S]
\Bigr).
\end{equation}
In general,
there are several distinct branches of stationary shapes, which we label
$S^{(n)}(v,\bar m_0)$, with corresponding energies $W^{(n)}(v,\bar m_0)$.
To
be a candidate for observation in the lab, a shape
$S^{(n)}(v,\bar m_0)$ must,
in addition, be locally stable to small shape perturbations
\cite{comment5}.
The lowest-energy branch (which must, of course, be
stable) defines the ground state and should, in principle, be observed at
sufficiently long times when the temperature is low.  However, when energy
barriers are large on the scale of $k_{B}T$, other low-lying locally stable
branches may remain metastable for long periods.  For SOPC, $\kappa \sim
20\ k_{B}T_{room}$ \cite{Evan87}, so metastability is expected
to
be common.

Finally, we shall need below an important connection between
the ADE-model shapes and those of the so-called spontaneous
curvature (SC) model\cite{Helf73,Miao91}, defined by the energy functional,
\begin{equation}
W_{SC}[S]= \frac{\kappa}{2}\oint
dA(2H-\bar {C_{0}})^2=\kappa \Bigl(G[S]-2\bar {c_0}m[S]+{\rm
const.}\Bigr),
\label{SCmodel}
\end{equation}
which describes a model without differential area elasticity
and having a spontaneous curvature $\bar C_0$ ($\bar c_0$ is
the corresponding reduced spontaneous curvature).  The variation
of Eq. (\ref{SCmodel}) gives a condition which has the same
form as Eq. (\ref{variation}) only with the replacement,
\begin{equation}
\label{effectivecurvature}
2\bar c_0 \equiv \alpha (\bar m_0-m[S^{(n)}]) .
\end{equation}
It follows that any stationary shape  $S^{(n)}$ of the ADE
model (\ref{ADE}) is also a stationary shape of a spontaneous
curvature model (\ref{SCmodel}) with the $\bar c_0$ defined by Eq.
(\ref{effectivecurvature}), which we shall henceforth refer
to as the effective reduced spontaneous curvature for the ADE
shape $S^{(n)}$. Notice that the control parameters
enter the variational shape equation (\ref{variation}) entirely
via the coefficient of the second term.
It follows that the stationary shape $S^{(n)}$ is specified completely and
in a way that is independent of $\alpha$ by
giving $v$ and $\bar c_0$.  For this reason, it
will sometimes be convenient in what follows to think of the stationary
shapes as $S^{(n)}(v,\bar c_0)$  rather than as
$S^{(n)}(v,\bar m_0)$, which still depends implicitly on
$\alpha$.  The variation (\ref{variation}) may be thought of as
proceeding in two steps:  First make $G[S]$ stationary at
fixed $m$, thus defining a function $G^{(n)}(v,m)$, then subsequently
carry out the variation with respect to $m$.
It follows from Eq. (\ref{variation}) that $\bar c_0$ can be
evaluated as
\begin{equation}
\label{curvature}
2\bar c_0=\frac{\partial G^{(n)}(v,m)}{\partial m}=2c_{0} + \alpha
(m_{0}-m[S^{(n)}]) .
\end{equation}

\subsection{$T=0$ Phase diagram and the stability of prolate shapes}
\label{phasediagram}

The map in the $(v, \bar m_0)$ plane of the regions where various
branches provide the lowest energy shape constitutes the $T=0$ (shape) phase
diagram for the ADE model.  This phase diagram, which depends on $\alpha$,
can
now be constructed rather easily numerically, at least, for $v$ not too
small and when
$\bar m_0/4\pi$
is not too far from unity
\cite{Miao94,Hein93,Jari95}.
The experiments described in this paper deal with a branch of axisymmetric
shapes, called ``prolates,'' because they have up/down symmetry and
resemble prolate ellipses when they have
reduced volume not too much below unity.  The region of the ADE
phase diagram in which these prolates appear is shown in
Fig.~\ref{fig:phasedia}.  It is
bounded
below by oblate axisymmetric shapes and by a region of non-axisymmetric
shapes, which need not concern us further here \cite{comment6}.
 Above the prolates lies a
region of pear shapes, for which axisymmetry remains but the up/down
symmetry
has been broken.  This pear region is, in turn, bounded above by a line
$L^{pear}$ of fully ``vesiculated'' limiting shapes, consisting of two
spheres
joined by a narrow neck. The region above $L^{pear}$ is
incompletely explored but
believed to be dominated by additional interconnected
vesiculated shapes.   The boundary between the prolates and pears
at relatively high reduced volume involves a
discontinuous shape change (corresponding to a simple crossing of the
energy
branches $W^{pear}$ and $W^{pro}$) along the line $D^{pro/pear}$
but a continuous shape change (corresponding to a
bifurcation of $W^{pear}$ away from $W^{pro}$) along the line labeled
$C^{pro/pear}$, for lower reduced volumes beyond
the tricritical point $T$. Both transitions are often called
``budding.''

It is important to emphasize that the prolate shape branch
continues to exist outside of the ``prolate region'' of the
phase diagram.  Indeed, within the context of the ADE model and
in the region of reduced volume shown
in Fig.~\ref{fig:phasedia}, a stationary prolate shape exists for every value
of $\bar m_0$
\cite{comment7}.
These shapes can only be observed, of course, when they are
locally stable.  The
region of {\it local} prolate stability includes the ``prolate
region'' of the phase diagram but extends beyond it into
metastable regions, where the true ground state has some other
shape.  It is a crucial test of the theory that prolate shapes
observed in the lab should, indeed, map to the region of
predicted prolate (local) stability.

 Metastability boundaries are marked by the first appearance of a
soft mode, i.e., a family of fluctuations which lower the
overall energy.  The region of Fig.~\ref{fig:phasedia} within which prolates
are predicted by the ADE model to be locally stable
is bounded above by the line $M^{pro}_{0,-}$ and
below by the line $M^{pro}_{2,+}$.  These lines are
calculated by an analysis of constrained Gaussian fluctuations
about the calculated stationary shape
\cite{Jari95}.  The subscripts label the rotation
mode $|m|$ and (even/odd) parity of the sector where the
first instability occurs.
It is an important result of this theory
\cite{Seif95}
that instabilities in sectors which break the symmetry are a
property of the shape $S^{(n)}$ alone, while those
in non-symmetry-breaking sectors depend in addition
independently on $\alpha$.  For the prolate shapes, the
boundaries $C^{pro/pear}$ and $M^{pro}_{0,-}$ in Fig.~\ref{fig:phasedia} both
reflect
instability in the symmetry-breaking sector $(|m|=0, {\rm odd\ parity\ })$.
These boundaries are, thus, independent of $\alpha$ in a
$(v,\bar c_0)$ representation of the phase diagram, so, in the
usual $(v,\bar m_0)$ representation, they shift with $\alpha$ according to
\begin{equation}
\bar m_0 = m[S^{pro}(v,\bar c_0)] +\frac{2\bar c_0}{\alpha}.
\label{mmapping}
\end{equation}
These lines of shape instability (and not the actual shape (phase)
boundaries!)
are the experimentally relevant
(observable) ones.

Strictly speaking, the above picture holds only in the
low-temperature limit, since for $T>0$ sharp phase boundaries do not exist,
because the vesicle is a finite system
with finite energy and always explores its full phase space.
Nevertheless, in practice, as long
as the prolate branch remains locally stable
and surrounded by energy barriers appreciably larger than
$k_{B}T$, thermally fluctuating prolate shapes are readily
seen in the lab.  In the prolate region of the phase diagram,
these fluctuating shapes constitute a true, stationary, equilibrium
ensemble.  In regions which are only metastable, the set of
prolate shapes should be
regarded as a restricted
ensemble, which may, however, be quasi-stationary for
appreciable periods of time.  In practice, metastability
is expected to break down slightly inside the boundaries $M^{pro}$,
when the metastability barrier becomes comparable to $k_{B}T$.

\subsection{Prolate shapes for reduced volumes near
unity:  The hierarchy and the mapping}
\label{mapping}

We review briefly here what is known theoretically
about the $T=0$ stationary
shapes $S^{pro}(v,\bar c_0)$ for the relatively high reduced
volumes which will be relevant for the experiments
\cite{Miao94,Jari95}.

Prolate shapes are axisymmetric. Therefore, they are
completely described by the curve
made by their intersection with any plane which includes the
symmetry axis.  This curve may be written in terms of an
arclength $s$ which starts at the north pole ($s=0$) and ends
at the south pole ($s=s^\ast$).  We take the direction of the
polar axis to be $\hat y$ and the perpendicular direction to be
$\hat x$. A representation which will be
convenient for our purposes is
 \begin{equation}
\psi(s) = \pi \frac{s}{s^\ast} + \sum_{n=1}^\infty a^{(0)}_n \sin(n \pi
\frac{s}{s^\ast})\ ,
\label{equ:expansionC4}
\end{equation}
where $\psi(s)$ is the angle between $\hat y$ and the
outward-pointing normal
to the curve. The overall length scale is set by $s^\ast$.
Note that $\psi(0) = 0$ and $\psi(s^{\ast}) = \pi$.  The first
term on the right describes a semicircular arc, i.e., a
spherical vesicle shape.  The coefficients $\{ a^{(0)}_n \}$ parameterize
deviations from the sphere.  For shapes like prolates, which are up-down
symmetric, the odd-$n$ coefficients vanish.  We note as an
aside that the coefficients $\{ a^{(0)}_n \}$ cannot be set
independently, since closure of the curve at the south pole
requires that
 \begin{equation}
x(s^{\ast}) = \int_{0}^{s^{\ast}}ds\,\cos \psi(s) = 0 \ .
\label{equ:condition}
\end{equation}
This places a complicated nonlinear condition on the set $\{
a^{(0)}_n \}$, which for any real vesicle shape will automatically be
satisfied.

The stationary shapes of the prolate branch are given by the
coefficients $a^{(0)}_{2n}(v,\bar c_0)$.  It is clear that, when $v$
is near unity, the coefficients ${a^{(0)}_{2n}}$ will all be small.  It
is a consequence of the stationarity condition
(\ref{variation}) that these non-vanishing
coefficients have the structure of a well-defined
hierarchy\cite{HGDthesis},
\begin{equation}
\begin{array}{lcccccc}
a^{(0)}_2\; &=& A_2\ (1-v)^\frac{1}{2} &+& B_{2}(\bar c_0) \ (1-v) &+& O(\
(1-v)^\frac{3}{2}\ )\ ,\\
a^{(0)}_4\; &=& & & \ B_{4}(\bar c_0)\ (1-v) &+& O(\
(1-v)^\frac{3}{2}\ )\ ,\\
a^{(0)}_6\; &=& & & & & O(\
(1-v)^\frac{3}{2}\ )\ ,\\
\end{array}
 \label{ourhierarchy}
\end{equation}
where $A_{2}=(135/64)^{1/2} \simeq 1.45$ and the coefficients
$B_{2}$ and $B_{4}$ are linear functions of $\bar c_0$ with coefficients
of order unity.
These results follow from Refs. \cite{Miao94} and
\cite{Miaothesis}. It is a consequence of this structure that for $v$
near unity $a^{(0)}_2 \gg a^{(0)}_4 \gg a^{(0)}_6 \dots$,
a hierarchy which reflects the fact that modes of higher
$n$ correspond to shorter wavelengths and cost more bending energy.
$a^{(0)}_2$ is independent of $\bar c_0$ at lowest order,
because this is the only contribution of order
$(1-v)^\frac{1}{2}$ and is, therefore, entirely determined by
the constraint on the reduced volume $v$.

Figure~\ref{fig:fish}, which was calculated by solving numerically the
variational
equation (\ref{variation}), illustrates the dependence of
$a^{(0)}_4$ on $v$ and $\bar c_0$ for the prolate branch near $v=1$.
It is clear that for $v<0.95$ the terms of order
$(1-v)^\frac{3}{2}$ and higher have an appreciable effect.
 Note that, except near the
anomalous point\cite{anomaly} $v\simeq 0.85$, knowledge of $a^{(0)}_4$ and
$v$ uniquely determines $\bar c_0$.

This brings us, finally, to the issue of the ``mapping'', i.e.,
of associating an
experimentally observed $T=0$ prolate vesicle shape with a point in the
ADE phase diagram, Fig.~\ref{fig:phasedia}.   Knowing the vesicle
shape means that we have direct experimental access to
``geometrical'' quantities such as $v$ and $m$, through  the shape
coefficients $a^{(0)}_{2n}$.  The abscissa, $v$, of the phase diagram
is geometrical; however, the ordinate, $\bar m_0$, Eq.
(\ref{m0bar}),
encodes information about the initial area difference $\Delta
A_0$ and the spontaneous curvature $C_0$, which are neither
geometric nor directly observable in any other way.  The
solution to this apparent impasse is to use $a_4$ and Fig.~\ref{fig:fish}
to infer a value of $\bar c_{0}(v,a^{(0)}_4)$, the effective reduced
spontaneous curvature, which is not
observable, and to combine this with the then (theoretically)
fixed
$m(v,\bar c_{0})$ to calculate $\bar m_0$
(Eq. (\ref{effectivecurvature})).  Note that, in principle, any of the nonzero
coefficients $\{a^{(0)}_{2n}\}$ could be used to produce such a
mapping, $\bar c_{0}(v,a^{(0)}_{2n})$.  In practice, however,
as the hierarchy (\ref{ourhierarchy}) shows,
$a^{(0)}_2$ is very insensitive to $\bar c_0$ (because $A_2$ is
independent of $\bar c_0$ and $B_2$ is only weakly dependent on
it) and $a^{(0)}_6$ is sufficiently small so that experimental
noise makes it a poor candidate.

This framework is still incomplete in three senses.  First, what the
experiment observes is not a single $T=0$ shape but an ensemble
of thermally fluctuating shapes.  Second, the mapping as
described above simply takes an experimental point and
associates it in a one-to-one manner with a point in the
theoretical phase diagram.  It does not yet in any obvious way test the
correspondence between theory and experiment.  Third, effects of
gravity should
be taken into account.  We discuss these
important points in Secs. \ref{shapecontours}, \ref{results}, and \ref{gravity},
 respectively.  But, before
doing so, we turn to the experiments.

\section{Experimental Techniques}
\label{techniques}

\subsection{Materials and Preparation}
\label{materials}

For all experiments, vesicles were prepared from the common phospholipids
1-Stearoyl-\-2-Oleoyl-sn-Glycero-\-3-Phosphat\-idyl\-choline (SOPC)  or
1,2-Dimyristoyl-sn-Glycero-3-Phosphat\-idyl\-choline (DMPC). These lipids
have
their main phase transitions at 5\cel\ and 23\cel, respectively
\cite{Mars90}.
They were purchased in powder form (Avanti Polar Lipids, Birmingham, AL, 
USA) and stored dissolved in
chloroform:methanol (2:1) in special chemically inert glass vials 
(Fisher Scientific)  below -15\cel.
 
Preparation was done using standard techniques
\cite{HGDthesis,Need88,Need93}: A few drops (30
$\mu$l) of lipid solution (10 mg per ml
chloroform:methanol) are spread with a syringe needle on a roughened Teflon
disk. The solvent is evaporated in a vacuum chamber overnight. The disk
with
the dried lipid is placed in a glass beaker (50 ml)  and pre-hydrated with
a
stream of Argon saturated with water vapor for about 20 minutes.
 Then, the desired solution for vesicle swelling is added and
the beaker is covered with Parafilm and placed in the oven. To avoid heat
shock, the solution and the beaker with the Teflon disk are heated
separately
to the swelling temperature prior to incubation. Swelling was done with 50
mMol
sucrose solution at a temperature of 36\cel.

Successful vesicle development is indicated by whitish streaks in the
swelling
solution. These streaks are collected into an Eppendorf tube
with cleaned glass pipettes and
incubated at the swelling temperature. Excess glucose solution
(48
mMol) is then added to obtain the desired  density
for the  vesicles in the observation chamber. The
end result of this procedure is a vesicle suspension with an interior
sucrose
solution and an exterior glucose solution (with a slight admixture of
sucrose).
The excess
 density of the interior relative to the exterior sugar solution is
 approximately 3.3 g/l.  This is needed in order that the vesicles sink
gently to
the bottom of the experimental cell, as discussed further below.
 Vesicles were stored at the swelling
temperature
 and used within a few days.

\subsection{Experimental Setup and Data Acquisition}
\label{setup}

For observation, vesicles are placed in a specially designed
microchamber, tightly sealed with glass plates above and below
to prevent evaporation \cite{HGDthesis}.  Temperature is
monitored by a thermocouple inserted into the observation
chamber.  A water bath,
incorporated integrally into the chamber, provides temperature
uniformity and
control at the level of $\pm 0.1$\cel. Because their density
is slightly higher than that of the surrounding solution, the
suspended vesicles fall to the bottom of the cell, where they
rest gently against the lower plate and are observed from below
via video phase-contrast
microscopy.  We use a standard inverted Leitz microscope
equipped with phase contrast (Leitz Phaco 40/0.65), capable of
an overall magnification of 500 times and incorporating annular
illumination and a phase ring.  The light source was a Hg arclamp
powered by a high-voltage transformer. A permanent green
filter and various grey filters
were used to minimize degradation of the lipids. The video camera was
positioned above the eye piece in such a way as to gain a resolution of 86
nm per pixel in a 480$\times$480 frame.

The visible phospholipid structures
\cite{comment8}
which collect at the bottom of the
observation chamber in the microscope's
focal plane are typically very diverse
\cite{Evan90,Kaes91},
including topologically
complex and multilamellar structures, small vesicles included
within larger ones, vesicles connected to one another by
sub-microscopic tethers or tubes, vesicles with obvious
adhesions, etc.  For detailed
observation, we try to select simple, topologically spherical,
unilamellar structures,
without identifiable microscopic connections or adhesions.
In addition, we monitor the fluctuations of
each candidate vesicle for some time prior to data
acquisition in order to reject those with obviously
``abnormal'' behavior, e.g., those exhibiting unexplainable
asymmetry or sudden changes in apparent area or volume.
The final fraction of usable vesicles is less than one percent.

Images of selected vesicles are simultaneously displayed on the
video monitor, saved to tape (U-Matic, Sony),
and processed in real time, as will be
described in the next subsection.
To sample at a fixed temperature a single thermal ``shape
ensemble'' takes about 20 min.  Recording a thermal shape
trajectory requires data at several different temperatures for
the same vesicle.  The chamber is allowed to equilibrate for
at least 5 min. after each temperature change.  Temperature is
recorded with a precision of $\pm 0.1$\cel. The total amount
of data gathered consists of  over 80 hours of video tape of more than 150
vesicles, including a wide range of shapes. Budding is an ubiquitous process,
which we observed at least 15 times in a {\em controlled} fashion.
It is important to get long runs at each
fixed temperature in order properly to sample the full thermal shape
ensemble.  On the other hand, thermal-trajectory runs which extend
over more than a few hours  appear to be contaminated by
systematic drifts, presumably due to lipid degradation and/or
flip-flop between bilayer leaves, which establishes intrinsic limits
on sampling density and run time.

In this paper we restrict
analysis to three particular SOPC vesicles (A, B, and C),
which have in common that they started with prolate
elliptical shapes and, on heating, eventually underwent budding
transitions, as illustrated in Figs.~\ref{fig:seq95}-\ref{fig:budding2}.
Qualitatively, other vesicles monitored behaved
similarly, although they followed different trajectories, some
exhibiting sharp shape transitions and others not.  The reason for
selecting the budding trajectories is that the location of the
budding instability provides a particularly stringent test of
the theory, as we shall discuss in Sec. \ref{results}.

\subsection{Processing the Video Image}
\label{videoimage}

In order to analyse the data, it is necessary to reduce the
video image to a time sequence of digitized shape contours.
This was done by using a frame grabber
(Matrox, Dorval, Quebec, Canada) to capture each image, computer processing the image in
real time, storing the digitized contour point in memory, and then
grabbing a new image.  The image processing algorithm
(described below) requires between  0.4 and 0.6 s (depending on
vesicle size) on a PC with a
 486 DX CPU and a 66 MHz clock speed.
Thus, for a video frequency of 30 frames/s, we are processing
every 15th frame.
This is relatively slow compared
to  processing times on the order of 0.1 s, which have been reported in the
literature \cite{Duwe90,Mele92}. However, in contrast to these fast
procedures,
our algorithm  has a better-than-pixel accuracy in finding the contour
\cite{Doug87,Doug88}. This high resolution turns out to be critical to
the success of our experiments, since we shall need to resolve
small changes in mean shape in tracking the thermal trajectory
\cite{comment9}.

In phase-contrast microscopy the image of the vesicle edge
exhibits a ``halo,'' with a light band (intensity maximum) just outside
the vesicle and a dark band (intensity minimum) just inside.  Typically, the
intensity profile crosses the gray value of the local background
at its steepest point, and we have taken this point to be the
nominal position of the vesicle boundary
\cite{Wils81}.

The contour-digitizing algorithm is fully described in Ref.
\cite{HGDthesis}.  The algorithm works on the (integer) pixel grid
($n_{x},n_{y}$) and
requires initialization by hand to the vicinity of the outline of the
particular vesicle to be studied (there are ordinarily several
vesicles in the field of view).  Suppose that a scan in $n_x$
at fixed $n_{y}=n_{y}^{(0)}$ intersects the halo.  By averaging
the grey values in the vicinity of the halo, we establish
a local background intensity.   The profile of
intensity-versus-$n_x$ crosses this background value at a point
$x$ (generally non-integer) which may be determined by linear
interpolation.  The point ($x,n_{y}^{(0)}$) is then stored as
a contour point, and the algorithm steps $n_{y}\to
n_{y}+1$ and starts again. Note that the interpolation
procedure allows determination of the $x$ coordinate of the contour point with
better-than-pixel precision.  Whenever the contour profile becomes
steeper when scanned in the $y$ direction rather than
in the $x$ direction, the
algorithm automatically switches to scanning $n_y$ at fixed
$n_x$, and vice versa. Each contour is terminated at closure.
 Motion of the
vesicle between successive frames is normally small enough so that each
subsequent frame can be started where the earlier one
terminated, so the initialization step needs to be carried out
only at the beginning of each run \cite{comment10}.

The digitized contours exhibit noise at the pixel level
(1 pixel=86 nm).  This behavior presumably reflects the intrinsic
noise of the original optical signal, the pixelation
statistics, the digitization of the grey scale, and other
factors.  To remove some of this microscopic
noise before data analysis, it is convenient to smooth the
observed contours.  This was done by applying a tenth-order binomial
filter \cite{filter} to the $x$ and $y$ contour coordinates,
thus averaging over an effective width of about 5 pixels.  The
distribution of deviations of the original data points from the
smoothed contour is Gaussian with a typical full width at half maximum
of about 0.7 pixels, thus giving an effective local lateral resolution of
about 30 nm (compared to a typical vesicle size of several
microns).  This resolution, well below the nominal
optical resolution given by the wavelength of light, illustrates
the delicate line-shape discrimination achievable via phase
contrast \cite{optics} and is more than adequate for quantifying the overall
vesicle shape and the low-lying fluctuation modes.

The result of this process is a time-sequence of several
thousand digitized contours, illustrating the
shape ensemble of each vesicle at each temperature.  The
relationship of these two-dimensional
contours to the three-dimensional vesicle
shape requires a brief discussion.
The ``general wisdom'' seems to be that what is seen in phase
contrast microscopy
is a cut through the vesicle in the focal plane
\cite{Duwe90,Mele92,Schn84,Enge85,Biva87,Fau89}.  However, this is
an oversimplification.  Phase contrast is particularly sensitive
to edges, so vesicle boundaries which
``overhang'' the focal plane (relative
to the optical axis) may contribute to
the image to a greater or lesser
extent depending on the focal depth and
the amplitude of the edge contrast.
Following the practice of the literature,
we shall ignore such effects in what follows.
 We wish only to point out that there are substantive issues here
 which deserve to be addressed more fully in future work.

In collecting data, the microscope is focussed on the
maximal cross section of the vesicle under observation,
and this focal plane does not change over time.  For vesicles such as
A, B, and C which are (on the average)
prolate and axisymmetric, gravity tends to orient the symmetry
axis horizontally, i.e., to bring it into the focal plane.
Thus, the contours (such as those based on 
Figs.~\ref{fig:seq95}--\ref{fig:budding2}),
which constitute our raw data, may be thought
of as an ensemble of cuts through the mean symmetry axis of the
fluctuating vesicle.  Fluctuations of the symmetry axis out of
the horizontal plane modify this simple picture:  If the
focal plane no longer includes the symmetry axis, then the
depth of focus and the edge enhancement mentioned in the previous
paragraph probably give the resulting image the character of a
projected outline of the tilted vesicle.  In principle,
this projected shape is
different from a true axial section.  In practice, the
stabilizing effect of gravity is large enough so that these
out-of-plane fluctuation effects are almost always small
 (except near the
spinodal line of the budding transition),
so we will treat
the two-dimensional contours as if they represent axial sections.
Note that there is a balance here.  In order to keep the theoretical
analysis simple, we would like to ignore the effects of gravity
on the vesicle shape.  On the other hand, in order to perform
the experiment conveniently, we use gravity to localize the vesicle in the
bottom of the chamber and to orient the symmetry axis (of
prolate vesicles) to the horizontal plane (see further
discussion at the end of Sec. \ref{results}).

\section{Analysis of the Two-Dimensional Shape Contours}
\label{shapecontours}

This Section describes how we parameterize the individual digitized
two-dimensional shape contours discussed in Sec.~\ref{techniques}, how we
average the shape parameters over each thermal
ensemble,
and how we infer $T=0$ three-dimensional
shape information from these averaged parameters.

\subsection{Thermal fluctuations of the vesicle shape:
General discussion}
\label{fluctuations}

Varying the temperature has two effects.  On the one hand, it
modifies, through ordinary thermal expansion, the
temperature-dependent control parameters, $A(T)$, $V(T)$, and $\Delta
A_{0}(T)$,
as well as the material control parameters, $\kappa$, $\bar \kappa$,
and $C_0$, which appear in the
Hamiltonian (\ref{ADE}).  These effects produce the so-called
``thermal trajectories,'' which we shall discuss in Sec.
\ref{results}.
On the other hand, even if all these parameters were
temperature independent, there would still be ordinary thermal
fluctuations.  It is for the moment these purely thermal
fluctuations to which we direct our attention.

There is, in principle, no way of taking a single
fluctuating shape contour and inferring the corresponding $T=0$
shape.  At best, we must take a full thermal shape ensemble and
use theory to infer the $T=0$ shape of the
vesicle {\it with the same control parameters}.  When the
fluctuations are large, even this is beyond present theoretical
capability.  However, when fluctuations are small enough
so that they may be treated at
the Gaussian level, progress can be made.

The upshot of a
recent study of
Gaussian fluctuations of vesicles of arbitrary axisymmetric
shapes \cite{Seif95,Jari95,Wort96} may be summarized as
follows:  Any typical fluctuating shape may be regarded as a
$T=0$ shape appropriately translated and rotated (the so-called
Euclidean modes) plus an area- and volume-conserving normal
(i.e., perpendicular) displacement $u({\bf r})$ at
each point ${\bf r}$ of the surface.
It is a special feature of these fluctuations
that (because of the strict area and volume constraints) both the
average displacement $\left<u({\bf r})\right>$ and the mean-square fluctuations
$\left<u^{2}(\bf r)\right>$ are generically of the order $k_{B}T/\omega$,
where $\omega$ is a typical static fluctuation-mode energy.  Note
that the
rms fluctuations are always larger than the shift when the
fluctuations are small.  In the analysis which follows, we
shall assume that the Gaussian regime holds and we shall ignore 
the mean shifts.    Ordinarily,
the fluctuation-mode energies $\omega$ are of the order $\kappa$.
For our vesicles
$\kappa/k_{B}T \simeq 20$, so for most regions of the phase
diagram, this is an excellent approximation.  There is,
however, an important exception.  At the instability
boundaries (Fig.~\ref{fig:phasedia}) one of the modes becomes soft.  Thus, near
enough such boundaries, the Gaussian treatment fails,
and we may expect difficulties (see Sec. \ref{results}).

\subsection{Parameterization of the two-dimensional shape
contours}
\label{parametrization}

We interpret the measured two-dimensional contours as being
sections which include the principal symmetry axis of the vesicle
(Sec. \ref{videoimage}).
Thus, in the spirit of the last paragraph and up to corrections
which are normally of order $k_{B}T/\kappa$, the center of mass
is located in (i.e., near) the focal plane at (i.e., near) the
point which is the center of mass of the digitized contour.  We
determine this point numerically for each contour.  We then
find the (approximate) principal (long) vesicle axis by calculating
 the two-dimensional moment-of-inertia tensor with
respect to the center of mass and
diagonalizing.  This determines the
(nominal) principal axis of each vesicle shape and identifies the
north and south poles.  We call the direction of the principal
axis $\hat y$ and the corresponding perpendicular direction $\hat x$
(which is, of course, not necessarily a principal axis of the
three dimensional vesicle).  Thus, each experimental contour is
reduced to a set of points $\{x_{i},y_{i}\}$.  In what follows,
we treat each half-contour separately (each image has two half
contours) and take $x_{i} \ge 0$.

It is convenient to represent each half-contour in the
angle-arclength $\psi (s)$ representation of Sec. \ref{mapping} by
calculating
\begin{equation}
\psi_i = -\arctan(\frac{y_{i+1} - y_{i-1}}{x_{i+1} - x_{i-1}})\ ,
\end{equation}
where the arctangent is defined on its Riemann surface,
i.e.,
$\psi(s)$ is continuous at the equator. The arclength $s$ is
measured from the north pole.
The parameterization
parallels Eq. (\ref{equ:expansionC4}),
\begin{equation}
\psi(s) = \pi \frac{s}{s^\ast} + \sum_{n=1}^\infty a_{n} \sin(n \pi
\frac{s}{s^\ast})\ .
\label{equ:expansion}
\end{equation}
The coefficients $\{ a_n\}$ are  obtained
by a
numerical integration using the trapezoidal rule,
\begin{equation} a_n = (-1)^n \frac{2}{n} + \sum_{i=1}^{M}  (\;\psi_i
\sin(n \pi
\frac{ s_i}{s^\ast}) +
\psi_{i+1} \sin(n \pi \frac{s_{i+1}}{s^\ast})\;)\
\frac{s_{i+1}-s_i}{s^\ast},
\label{equ:expamp}
\end{equation}
and, henceforth, they replace the points
$\{ x_{i},y_{i}  \}$ in representing the
half-contour.  ($M$ is the number of digitized points in the
half-contour.)  Note that the contours here are not
up-down symmetric, so the odd-n coefficients do not in general
vanish, as they did for the $T=0$ prolate shapes.
Similarly, $\psi (0)$ and $\psi ({s^\ast})$ are normally nonzero.

\subsection{Thermal ensembles and $T=0$ shapes}
\label{shapes}

For each half-contour of each video image, we calculate the
shape coefficients $\{a_n\}$ plus the nominal (``effective'')
vesicle area and volume,
\begin{equation}
\label{equ:area}
 A_e  =  \pi  \sum_{i=1}^{M} (x_{i} + x_{i+1}) (s_{i+1}-s_i)
\end{equation} and
\begin{equation}
\label{equ:volume}
 V_e =  -\frac{\pi}{2} \sum_{i=1}^{M} (x_i^2 + x_{i+1}^2) (y_{i+1} - y_i) .
\end{equation}
In a similar spirit, we compute an effective reduced
volume,
\begin{equation} v_e =
\frac{V_e}{\frac{4\pi}{3}(\frac{A_e}{4\pi})^\frac{3}{2}}\ .
\label{equ:effvol}
\end{equation}
For axisymmetric vesicle shapes, these equations would calculate
the true area, volume, and reduced volume, respectively.  Since
each image is only a snapshot of a section through a
fluctuating vesicle, $v_e$ is only approximately equal to the
true reduced volume $v$.  Note that $v_e$ fluctuates in time for
successive images of a given vesicle at a fixed temperature,
while $v$ is in principle constant, since the true area and volume
are conserved during the shape fluctuation.

The several thousand images which constitute a typical
experimental run with a given vesicle at a fixed
temperature lead to characteristic
time series for the quantities $\{ a_n \}$ and $v_e$,
as illustrated by Figs.~\ref{fig:tseries1} and \ref{fig:tseries2}.  
Although the series are noisy,
we expect to see memory effects
between successive images,
as long as there are any characteristic physical
relaxation times longer than
the 0.5 s between successive grabbed images.  We have estimated
elsewhere \cite{Doeb95} the typical
relaxation times expected for these
vesicles.  Away from instabilities, the characteristic times are
expected to be at most several seconds, which is consistent
with direct visual observations of the optical image.  As the vesicle
approaches an
instability (which occured at $v=0.878$ for vesicle A), one sees
rapidly increasing relaxation times, corresponding to a
spinodal slowing-down\cite{Doeb95}.  This tendency is clearly
visible when comparing Figs.~\ref{fig:tseries1} and \ref{fig:tseries2}.
As long as the data set spans a time interval much larger than
the longest relaxation time, we may expect that the time
sequence samples an effective stationary
ensemble \cite{comment12}.  In this
sense, the thermal ensemble of fluctuating shapes is
characterized by the set of ensemble averages $\left<a_n\right>$,
$\left<a_{m}a_{n}\right>$, $\left<v_e\right>$, $\left<v_{e}^{2}\right>$, etc.  Indeed, the
distribution functions $P(a_{n})$, $P(v_{e})$, etc., are typically
Gaussian form in shape \cite{HGDthesis}.

In order to proceed with the mapping, we now need a procedure
for inferring the zero-temperature quantities $v$ and
$\{a^{(0)}_{n}\}$ from the thermal data.  We do this in the crudest
way, by simply making the identifications,
\begin{equation}
v=\left<v_e\right> {\rm\ and\ } a^{(0)}_{n}=\left<a_{n}\right>.
\label{equ:ident}
\end{equation}
As explained above (Sec.\ref{fluctuations}),
the justification for these
identifications is that the averaging process suppresses the rms
fluctuations, which are of order $\sqrt{k_{B}T/\omega}$.  This
leaves the thermal shifts plus the terms of order $\left<u^2\right>$, both
of which scale as $k_{B}T/\omega$, which
we ignore in first approximation\cite{comment13}.  These values
of $v$ and $a^{(0)}_4$ allow us (Sec. \ref{mapping})
to infer $\bar c_0$ (and, thereby,
$\bar m_0$) from Fig.~\ref{fig:fish}, and, thus, to complete the mapping.

It is hard to give any meaningful estimate of the real
uncertainty in the derived quantities $v$ and $a^{(0)}_n$.  For a
truly stationary ensemble, the purely statistical (sampling)
uncertainties in the average quantities should decrease as the
ensemble sampling becomes denser.  In practice, our runs are
necessarily of finite length (Sec. \ref{setup}).  Indeed, if we divide
the data set into two parts, corresponding to earlier and later
times, we typically see a spread of values
corresponding typically to $\pm\ 0.001$ for $v$ and $\pm\ 0.002$ for the
$\{ a^{(0)}_n \}$'s (and somewhat larger near the budding instability).
It is this measure  which we adopt
as an estimate of the statistical  uncertainties.
Of course, there are also
systematic errors, such as the thermal shifts (which we have neglected), the
failure of the Gaussian picture (where fluctuations are large), the sampling
error  (when the relaxation times are long), the fluctuations of the major axis
out of the focal plane, and the effects of gravity (Sec.
\ref{gravity}).   The upshot is
that well away from the instability boundary the statistical uncertainties are
probably realistic, except for the systematic influence of gravitational
effects.  Near the instabilities, the situation is less well defined.  These
statistical uncertainties translate (via Fig.~\ref{fig:fish}) into
uncertainties in
$\bar c_0$, as we shall illustrate in Sec.~\ref{results}.

\section{Results (without gravitational corrections)}
\label{results}

Each of the three budding vesicles, A,B, and C, started at a
relatively low temperature with a nearly spherical shape
(i.e., $v \approx 1$).  As the temperature was raised, the
reduced volume decreased, until at a certain temperature
(different for the different vesicles) a ``budding'' instability occurred
(see Figs.~\ref{fig:budding} and~\ref{fig:budding2}), i.e., the vesicle suddenly necked down and,
over a time interval of $1-10$ seconds, developed
a small quasispherical satellite.  (This time range is due to the
different vesicle sizes, since typical relaxation times scale with the third
power of the vesicle radius \cite{Doeb95}.)  Up to the budding
threshold, the thermally induced changes in the fluctuating
ensemble are reversible to within experimental precision.  The
budding process, itself, is a mechanical instability
\cite{Doeb95}.  In fact, the budding can be reversed, but only by
cooling to a temperature significantly below the budding
temperature\cite{comment14}.
The size of the fluctuations and the scale of the longest
relaxation time increases dramatically as the temperature
approaches the budding temperature (see Figs.~\ref{fig:seq95},
\ref{fig:seq91}, \ref{fig:tseries1}, and \ref{fig:tseries2}).   These effects have been
interpreted in terms of a simple Landau theory\cite{Doeb95}.

Our results for the three budding vesicles, A, B, and C, are
summarized in Table~\ref{table}.
The average amplitudes $\left<a_n\right>$ were generally very small for odd 
$n$, as expected in the prolate phase\cite{comment15}.
The even coefficients $\left<a_6\right>$ and above were too small to
distinguish from zero, presumably because of the hierarchy
(\ref{ourhierarchy}).
The value $v_b$ of the reduced volume at budding was determined
by extrapolating the experimental temperature dependence $v(T)$ to
the observed budding temperature $T_b$.

Figure ~\ref{fig:cmap} shows the result of
mapping this data into the theoretical $(v,\bar c_0)$ 
diagram by using $\left<v_e\right>$, $\left<a_4\right>$, and Fig.~\ref{fig:fish},
as explained in Sec. \ref{mapping}.  The instability lines $M^{pear}$ and
$M^{obl}$ are just the appropriately mapped versions of the
corresponding spinodal lines of Fig.~\ref{fig:phasedia}.  The
advantage of this representation is that it is completely
independent of the value of $\alpha$, as explained after Eq.
(\ref{effectivecurvature}).
Theory predicts that the prolate shapes are locally
stable only between the two spinodals.  With the exception of
the highest-temperature point in the trajectory of vesicle A,
we see that the mapped shapes do lie in this region.  Fig.~\ref{fig:mmap}
 shows the same data plotted in the $(v,\bar m_0)$
phase diagram, Fig.~\ref{fig:phasedia}.
The required relation between
$\bar c_0$ and $\bar m_0$ is based on
Eq.~(\ref{effectivecurvature}). In order to evaluate $m[S^{(n)}]$, we solve
the variational shape equations derived from energy functional 
Eq.~(\ref{SCmodel}) for the given values of $v$ and $\bar {c_0}$.
We have taken $\alpha=1.4$ in making this transformation \cite{Waug92,Miao91}.
Since the spinodal boundaries map right along with the data
points, there is no change in the predicted stability.

The fact that, with a single exception, the mapped data points
lie neatly sandwiched in the region of
predicted (local) stability is a
stringent quantitative test of the theory and constitutes the
single most important result of this work.  We emphasize again
(see Sec. \ref{phasediagram}) that the shapes mapped by
Fig.~\ref{fig:fish} are
variationally stationary (by construction) but {\it not} necessarily locally
stable, so that an arbitrary shape could end up anywhere in the
phase diagram.

It is worth pointing out that fact that the values
of $\bar c_0$ derived from the data points are all of order
unity (as expected on the basis of the theory) is also an
important test.  Fig.~\ref{fig:fish} shows that $\bar c_0$ values
between $-5$ and $10$  are associated with values of $a^{(0)}_4$
in the narrow range between $-0.02$ and $0.02$.  If the theory
were significantly in error, it would be quite easy to have
produced very large or very small values of $\bar c_0$ .

Indeed, in a certain sense, our ability reliably to distinguish
shape changes corresponding to differences of order unity in
$\bar c_0$ is, in itself, surprising.  Consider that, for a
vesicle of radius 10 $\mu$m at a reduced volume $v=0.9$, a
difference in $\bar c_0$ of $\pm 1$ corresponds to a change in
shape which modifies the pole-to-pole contour length $s^{\ast}$
by only 20 nm.  This number (the smallness of which is a direct
consequence of the hierarchy) is below the
{\em local} lateral resolution of
the contour.  How is this possible?  First, one has to
realize that one does not measure a single distance only. Rather, the
amplitudes
 are calculated {\em globally} from an integral
 (see Eq.~(\ref{equ:expamp}))  over about 600
 contour points, each of which deviates from the reference shape.
Second, one is interested in a low mode,
which is insensitive to local
perturbations in the membrane. And, third,
the amplitudes are averaged over
typically  several thousand contours, giving an effective sample
size on the order of $10^5$. Thus, shape differences
on the $10\ $nm scale
are detectable \cite{optics}.

The ``thermal trajectories'' corresponding to each vesicle
encode the effect of the experimental control
parameter (temperature) on the quantities $v$ and $\bar c_0$
(or $\bar m_0$), defined in Sec. \ref{ADEmodel}.  These quantities, in
turn, depend on the volume $V$, area $A$, relaxed area
difference $\Delta A_0$ of the vesicle, on the thickness
$D$ and spontaneous curvature $C_0$ of the membrane, and on the
ratio $\alpha$ of elastic constants.  All these quantities are
in principle temperature dependent, and, if these dependences
were known, we could calculate the thermal trajectory and
compare with that found in Figs.~\ref{fig:cmap} and~\ref{fig:mmap}.\
  The volume thermal
expansion coefficient
($\beta _{V}\approx 3\times 10^{-4}$/K for water) is known
to be small compared to the area thermal expansion coefficient
($\beta _{A}\approx 3\times 10^{-3}$/K for SOPC\cite{thermalexpansion}).  It is
also known \cite{volcons} that the total bilayer volume
$AD$ is only weakly temperature dependent.  A simple
model is to assume that only $A$ and $D$ are temperature
dependent.  When $C_0=0$, as is reasonable for a symmetric
bilayer, this assumption leads to the simple result that the
product $v\bar m_0$ is temperature
independent\cite{Seif91,HGDthesis}.
This hypothesis predicts trajectories of the general shape
and scale
shown in Figs.~\ref{fig:cmap} and~\ref{fig:mmap} but significantly less steep
than
those observed.  It is not hard to make more refined models
consistent with the data, for example, by using a
non-zero spontaneous curvature $C_0$ and/or a differential
thermal expansion for the two leaves of the bilayer
\cite{Seif91,HGDthesis}.  Unfortunately, direct measurements of
these quantities are not available, so no useful conclusions
can be drawn at this stage.

Another set of evidence bearing on the consistency of the
observations with the theory is the relative size and shape of the main
vesicle and the bud which forms at the instability. At the
crudest level, theory predicts that the final state after
budding will be pear shaped (rather than fully vesiculated,
with a microscopically narrow neck) when the budding occurs
for reduced volumes less than $v_c=0.875$,
where the spinodal crosses the limiting-pear line $L^{pear}$
(see Fig.~\ref{fig:phasedia}).
As $\alpha$ increases, this crossing point moves to higher
values of $v$, so the observation that vesicle A buds to a
fully vesiculated state would be inconsistent with a value of
$\alpha$ larger than 1.4.  This observation places an upper bound on 
$\alpha$.
 Another qualitative
prediction is that the ratio $r$ of the bud radius to the radius of
the remaining main vesicle should increase as the reduced
volume at budding
decreases. Thus, $r$ should be largest for vesicle A and smallest for
vesicle C, as is, indeed, observed.  On the other hand, on the
limiting line $L^{pear}$, the vesiculated configuration
consists of two spheres, so there is a
simple relation between  the ratio $r$ and reduced volume  $v$
\cite{Miao91,comment16}.
In particular, close examination of Fig.~\ref{fig:budding} provides a value of $r$ close to 2.8 (pure geometry, since the shape is very close to
being two spheres). This corresponds to $v(r)=0.873$. Indeed, budding is observed at $v_b=0.878\pm0.002$ for vesicle A.

Overall, the agreement of theory and
experiment is reasonable with the exception of the last point of
the thermal trajectory of vesicle A, which lies distinctly
above the theoretical spinodal boundary $M^{pro}$ (Figs.~\ref{fig:cmap}
and~\ref{fig:mmap}).
This last point is worrying.
Indeed, even below the
spinodal line, there should be a (fuzzy) unstable region where
the energy barrier out of the metastable state is of order
$k_{B}T$.  (The fact that vesicles B and C
appear to bud increasingly below $M^{pro}$ for higher reduced
volume suggests that there may be some systematic effect at
work which is distorting the locus of instability.)  We have
considered three possible reasons for this discrepancy.

First, our identification (\ref{equ:ident}) involves the
assumptions that (a) the fluctuations are small enough to be
treated at
the Gaussian level and (b) the (Gaussian) thermal shifts  and
rms fluctuations (of order
$k_{B}T/\omega$) can be neglected.  At the spinodal, fluctuations
diverge\cite{Doeb95}, so neither of these assumptions is valid,
and the identification (\ref{equ:ident}) is expected to fail.
It is entirely plausible that these assumptions are already
breaking down near the spinodal, at the last stable point.
Because the effects of fluctuations beyond the Gaussian level have yet
to be calculated, we cannot at this stage assess the impact that such 
corrections might have on the near-spinodal points of trajectory A 

Second, we have assumed that the major prolate axis is
(effectively) in the
focal plane of the microscope.  If this axis is appreciably out
of the focal plane, then the digitized images cannot be thought
of as sampling axial sections of the three-dimensional
fluctuating shape, and the whole analysis of these images would
have to be redone.  As long as fluctuations are small, it is
reasonable to assume that gravity acts to keep the prolate axis
aligned.  But, near the spinodal line, there are large, slow
pear-like fluctuations\cite{Doeb95}, which are not
``up-down'' symmetric (i.e., which break the symmetry between the north
and south poles).   In this situation, gravity may be expected to
systematically re-orient the small end of the pear towards the
bottom of the chamber, thus tipping the effective symmetry
axis away from the horizontal.  Once tipped, the
symmetry axis is  inhibited by gravity from returning to
the horizontal, so one expects long intervals of
asymmetric, pear-like data to appear in the near-spinodal
time-sequences.  In fact, the data for the last point of the
vesicle A trajectory do show an ``anomalous'' cluster of frames with
simultaneously large $a_3$ and $a_4$, and these frames exhibit a fuzzy contour
profile near  the small end of the pear, indicating an overhang of the vesicle
membrane beyond the focal plane.
  The effect of excluding this segment of the time-sequence is
to lower $\left<a_4\right>$ nearly to the spinodal, thereby improving agreement
between theory and experiment.

Finally, we have so far treated the effect of gravity as only
something which positions the vesicles at the bottom of the
chamber and aligns (prolate) axes in the horizontal plane.  In
fact, it will also modify the zero-gravity shape analysis which
has been up to this point the basis of our
mapping procedure.  What effect do
gravitational shape changes have on the analysis and can they
explain the observed discrepancies?
This is the subject of
the next Section.

\section{Gravitational Effects}
\label{gravity}

\subsection{Qualitative considerations}
\label{qualitative}

When the density of the solution which fills the vesicle
interior is greater than that of the exterior solvent, the
vesicle will fall to the bottom of the container and, once in
contact with the bottom, will deform in such a way as to
decrease the gravitational potential energy of the interior,
higher-density material.  The overall shape involves a balance
between the previous bending energy (\ref{ADE}) and a new
gravitational energy,
\begin{equation}
  W_{grav}[S] =  g_0 \Delta \rho \int z\,dV\,,
  \label{free_en}
\end{equation}
where $g_0$ is the local acceleration due to gravity, $\Delta
\rho$ is the excess mass density of the interior solution, $z$
measures height above the bottom of the chamber, and the
integral is over the interior volume of the vesicle.

The ratio
of the energy scale $g_0 \Delta \rho R_A^4$  of this gravitational term to
the scale $\kappa$ of the bending energy defines a
dimensionless parameter,
\begin{equation}
g \equiv  \frac{g_0\,\Delta\rho\,R_A^4}{\kappa}\,,
  \label{gquer}
\end{equation}
which measures the relative size of gravitational and bending
energies\cite{Krau95}.  When $g$ is very small, we may expect shapes
which are not significantly deformed relative to the
gravity-free case.  When $g$ is very large, gravitational energy
dominates and vesicles will tend towards circular pancakes \cite{Krausthesis},
squashed against the chamber bottom, insofar as constraints on
area and volume allow.  (Of course, if $v=1$, then the vesicle
can only be spherical.)

Experimental values for our vesicles
A, B, and C were nominally $g=$2.2, 0.3, and 1.9,
respectively\cite{comment17}.  Here, we use a value of
$\kappa = 0.9 \times 10^{-19}\  J\ $ for our estimation \cite{Evan90}.
 We may, thus, expect gravitational
corrections to be appreciable for vesicles A and C but
relatively less important for vesicle B.  The qualitative
effect of gravity on the mapping is not hard to see.  Roughly
speaking, a ``pancake'' deformation will make the focal-plane
section of a prolate rounder and larger in area than it would
otherwise be.  Therefore, gravitational corrections
will lead to larger values of $v_e$.
Although the coefficients $a_n$
must approach zero for large g, it is not obvious where the
asymptotic regime sets in, so the sign of the gravitational shift in
$a_4$ cannot be inferred a priori.  In order
to estimate these effects quantitatively, we need to be able to
calculate vesicle shapes in the presence of gravity.

\subsection{$T=0$ shapes in the presence of gravity}
\label{gravitationalshapes}

The only previous calculation of vesicle shapes including
gravitational effects was done by Kraus et al. \cite{Krau95}.  These
authors found a gravity-induced prolate-oblate transition for values
of $g$ similar to those encountered in our experiments. Following this
work, we use a polyhedral discretization of the vesicle surface and
employ the program {\tt Surface Evolver} \cite{Brak92} to search
iteratively for the shape-energy minimum.  Numerical minimization in
the presence of a hard-wall constraint for the chamber floor leads to
special problems in stability.  For this reason, we replaced the hard
wall by a soft substrate potential, $V_{\rm w}(z) = V_{0\rm w} \exp(-z
/ z_0)$, with $V_{0\rm w}=5 \kappa$ and $z_0 = 0.1 R_A$.  These
parameters seem to provide a good compromise between numerical
stability (favored by a softer potential) and a deformation of the
shape caused by the soft tail of the potential which is as small as
possible \cite{Krau95}.  Constraints on area and volume are respected.
The energy is minimized by moving the vertices in the direction of the
energy gradient or, alternatively, by a conjugate gradient method.
Symmetries such as mirror planes can be exploited. For most of this
work, only a vertically-cut quarter section of the vesicle was
actually computed. It turns out that the results for the final shape
and energy are very sensitively dependent on the triangulation in a
way that we cannot completely control.  We have tried to
overcome this problem by fitting a linear
interpolation to a grid of data points,
as described in Sec. VI B below.  This procedure averages out
random fluctuations from one point to another but cannot address
any subtle systematic dependence on grid size which might be hidden beneath
the fluctuations (we did check directly for such a grid-size effect, and none is
apparent at the level of accuracy we can achieve).  In the absence of a more
reliable measure, we have simply used the deviation of the computed data points
from the smoothed interpolation to give an estimate of the error introduced
by the
triangulation.

Finally, we point out that, not only does gravity influence the
shape and energy of a vesicle at given $v$ and $\bar c_0$, but
 it also changes the relative energy of different shape branches,
thus shifting phase boundaries and stability boundaries in the
phase diagram.  Thus, in looking for gravitational corrections
to the experiments, we also need to compute the gravitationally
shifted spinodal line $M^{pro}$.  We have done this
in the ($v,\bar c_0$) representation (Sec. \ref{ADEmodel}) for $g=2.2$,
 in which now
(c.f., Eq.(\ref{curvature}))
\begin{equation}
\label{equ:curvature}
2\bar c_0=\frac{\partial G(v,m)}{\partial m}+
\frac{1}{\kappa}\frac{\partial W_{grav}(v,m)}{\partial m}=2c_{0} + \alpha
(m_{0}-m[S^{(n)}]) .
\end{equation}
This involves two numerical fits to {\tt Surface Evolver} data,
first for the computation of $m(M^{pro})$ and then for the derivative which
evaluates $\bar{c}_0$. As a consequence, the quality of the results is rather
poor.
As shown in Fig.~\ref{fig:cmap},
the  result for our experiments is a shift of the upper spinodal
upward in $\bar{c}_0$ by about one unit; but, the
numerical uncertainties are unfortunately comparable in size to
the shift.  More detailed calculations would require finer
triangulations and much longer relaxation times.  Since the
computational investment is already
appreciable and experimental
uncertainties are already large near the spinodal, additional
investment at this time does not seem wise.
We did not compute the location of the full lower spinodal including gravity;
however, for $\bar c_0=0$ (and $g=2.2$), we do know \cite{Krau95} that the
prolate-oblate transition occurs at $v=0.94$ (as plotted), which also
corresponds
to a small upward shift.

\subsection{Gravitational corrections to the mapping}
\label{corrections}

We estimated gravitational effects by running the {\tt Surface
Evolver} program for $v$ between 0.875 and 0.975 in steps of 0.025 and
for $\bar c_0$ between 2.5 and 10.0 in steps of 2.5. For each pair of
these parameters, we computed shapes over a range of small $g$ values.  For
each shape, we took a maximal horizontal section and computed effective
values of $v_e$, $a_2$, and $a_4$ using formulas (\ref{equ:expamp}) and
(\ref{equ:effvol}).  These values varied in a roughly linear way with
$g$, only with some superimposed fluctuations which we attributed to
the triangulation sensitivity mentioned in the previous subsection.
We then assigned effective values of $v_{e}(g)$, $a_2(g)$, and $a_{4}(g)$ for the
gravitationally distorted shapes by making a straight-line
fit to these computed points
passing through the values previously computed for $g=0$.
For the values of $g$, $v$,
and $\bar c_0$ relevant to the experiments, the gravitational shift of
$a_4$ is comparable to the numerical errors. On the other hand, the difference
between $v_{e}(g)$ and the actual reduced volume $v$
 (see Fig.~\ref{fig:veff}) does lead
to a noticeable correction in the values of $\bar c_0$ inferred from the
experimental data for vesicles A and C.

This family of
lines may then be used to calculate
 gravitational corrections to the experimental data.
Since $g$ is known, it is only necessary to take the measured values of
$v_e$ and
$a_4$, which belong (presumably) to shapes which are gravitationally distorted,
and to infer the corresponding values of $v$ and $\bar c_0$.
Figure \ref{fig:veff} shows, for example,
the calculated gravitational shift in the
apparent volume, ($v_{e}-v$), at $g=2.2$ (appropriate for vesicle A) for
representative values of $v$ and $\bar c_0$.
Note that the $v_e$ is always larger than
$v$, in agreement with the qualitative argument of Sec. VI A, so that
gravitational correction always shifts the data points to the left in
Fig.~\ref{fig:cmap}.
 The corrections increase for values of
$\bar c_0$ close to the prolate-oblate transition, i.e.,
for small  $\bar c_0$, where increasing the volume of a prolate vesicle at
fixed $\bar c_0$ leads eventually to a transition to an oblate
shape with a vertical symmetry axis, thus producing a circular
focal-plane section (i.e., $v_e = 1$).  The behavior of the shift
of the apparent volume near this prolate-oblate transition can be understood as
follows: Above the transition, for $v > v_c(\bar c_0,g)$, the focal cut of the
oblate vesicle
is circular, independent of $( c_0,g)$, i.e.,
\begin{equation}
 \label{equ:rtrans}
v_e - v = 1-v\ ,\ \   v > v_c
\end{equation}
 For $v < v_c$, we find from geometry $v_e = 1 - \frac{64}{135} a^2_2\ +\ O(
a^4_2, a^2_4,\cdots)$.  Furthermore, the amplitude $a_2$
 is given by $a_2 \approx c
(v_c-v)^\frac{1}{2}$, where the coefficient
$c$ depends on both $\bar
c_0$ and $g$. Thus, immediately below the transition, we have
\begin{equation}
  \label{equ:ltrans}
  v_e -v \simeq 1 - \frac{64}{135} c^2 v_c + ( \frac{64}{135} c^2-1)\ v\ ,\ \
  v \lsim v_c\ .
\end{equation}
This equation implies that all the shifts in the apparent reduced volume for
different $\bar c_0$ meet the curve $v_e-v = 1 - v$ with a slope,
$( \frac{64}{135} c^2-1)$, larger than $-1$.
Inspection of Fig.~\ref{fig:veff} suggests that the slope is, in fact,
positive for small  $\bar c_0$.

Figures~\ref{fig:cmap} and \ref{fig:mmap} show
the gravitationally corrected (phase)
diagrams with the corrected data points for vesicles A and C. The uncertainties
of the gravitationally corrected points include both the original
experimental uncertainties and the numerical uncertainties of the
gravitational shape-energy calculations.  Note that the data points are
shifted to the left in $v$, as expected qualitatively.
All experimental points are in the (metastable) prolate phase,
except the ``bad'' point which still remains above the upper spinodal,
unless the ``anomalous'' cluster is removed, as discussed in the last
section. Accepting this somewhat ad hoc procedure, one may argue that
this last stable point at $v = 0.890$ lies within error bars inside the prolate
phase after the gravitational corrections have been performed.  We note,
however,
that the actual point of budding at $v_{b}=0.878$, which must in principle be
beneath the spinodal, would still appear to lie slightly in  the unstable
region,
even after gravitational and tipping corrections.  We may speculate that this
apparent inconsistency is due to thermal shifts (neglected so far in our
treatment), which could be appreciable near the spinodal.

So far, we have discussed stability of vesicle $A$ only with respect to the
upper (pear-mode) spinodal.  We have also checked that the data points of
vesicle
$A$ fall above the lower limit of stability of the prolates, which is an
instability towards the oblate phase.  In fact, the first point of
vesicle $A$ is located (including the gravity correction) at $(v=0.937,\bar
c_0=1.6)$. The location of the prolate-oblate
transition at the same volume (and $g=2.2$) is known to occur at $\bar
c_0 =0$ (see the previous section), which is comfortably below our data point.
Thus, the vesicle-A trajectory does (properly) start in the stable-prolate
region. On the other hand, this first point still appears somewhat out of line
with the remaining three points of the trajectory (see Fig.~\ref{fig:mmap}),
which
(after gravity corrections) fit quite well to the simple form
\cite{Miao94}, $\bar
m_0 v=const$.  We may speculate that thermal shifts play a role here, too, near
the lower spinodal \cite{comment18}.

A few comments are in order concerning gravitational corrections to
the vesicle-C data.  The reduced volume of vesicle $C$ is shifted to
the left, as expected.  Unfortunately, the effective spontaneous
curvature $\bar c_0$ is not well determined, due to the large
experimental and numerical errors. (We remind the reader that vesicle
$C$ did bud at this location and, thus, exhibited large spinodal
fluctuations.)  This data point appears to be located appreciably
below the spinodal line. This could be an artifact created by effects
not included in the mapping (see the discussion at the end of
Sec.~\ref{results}) and/or it could be due to a low activation energy
for budding near the sphere.

The upshot of this exploration of gravitational corrections is
that gravity does, indeed, have a substantial numerical effect, as
might be anticipated from the fact that the dimensionless parameter
$g$ is around $2$ for vesicles A and C.  However, the
qualitative (and generally encouraging) conclusions of the
gravity-free analysis are not changed.

\section{Conclusion}
\label{conclusion}

Previous experiments (e.g., Ref \cite{Bern90}) have compared
experiment with theory by, in effect, exhibiting a set of control
parameters ($v$, $c_0$, $m_0$, $\alpha$) which lead to theoretical
shapes similar to those observed in the laboratory.  It is important
that this exercise can be successfully carried through; but, for
various reasons, it constitutes far less than a full test of the
theory.

The first problem is that different variants of the theory,
ranging from the SC model ($\alpha =0$) to the $\Delta A$
(bilayer-couple) model ($\alpha =\infty$),
all share the same set of
stationary shapes, so that observation of a shape which can be
suitably parameterized only distinguishes models in which the
shape is stable from those in which it is not.  One would
like to be able to measure all the
control parameters for a given vesicle and then to verify that
a vesicle with those control parameters does, indeed, have the
observed shape.  The difficulty is that, while $v$ and $\alpha$
are measurable, $c_0$ and $m_0$ (which enter the shape problem
in the combination $\bar m_0$) are not.  We have surmounted
this problem by concentrating on the equivalent variable $\bar
c_0$ (which incorporates $c_0$, $m_0$, and $\alpha$) and
inferring this variable directly from the shape data ($a_4$).
Although the inference process uses theory, there are
nontrivial checks left over.  Local stability is still an
important check, as we have argued.  In particular, the
observed (reasonable) agreement of the  experimental
budding boundary with the calculated theoretical spinodal is
encouraging, as is the qualitative agreement of the
post-budding shape with that predicted by the theory.
Once the mapping is done, observation of other shape
coefficients ($a_2$, $a_6$, etc., for the prolates)
provides, in principle, a further test of
agreement between theory and experiment.
 Unfortunately, at the level of precision we have
been able to achieve here, $a_2$ is too weakly dependent on
$\bar c_0$ to be useful, and $a_6$ is too small.

The second and in many ways more important advance which our
experiment makes over previous ones is in the monitoring and
analysis of the full thermal shape ensemble.  Previous workers
have certainly observed the shape fluctuations; however, shape
comparisons between theory and experiment have heretofore
relied on comparison of a single judiciously selected image with a
theoretically calculated shape.  When fluctuations are
appreciable (which they certainly become near any instability
boundary), this process is clearly unacceptable.  We have
illustrated how to monitor and to analyse the full shape
ensemble, and we have shown how to relate the ensemble data to
the corresponding $T=0$ theoretical shapes, at least in
situations where fluctuations are not too large.  Treatment of
larger fluctuations, which are common near instabilities
and will certainly be increasingly important at low $v$ (where
mechanical modes will tend to be softer), will require a new
theoretical approach capable of going beyond the Gaussian
level.

Finally, at a somewhat technical level, we have illustrated
that the effects of gravity, which have been ignored in earlier
work, are quantitatively important.  And, we have shown how to
adjust for them in comparing theory and experiment.

In summary, our work provides in principle a quantitative test of
the ADE model of vesicle shapes.  Agreement between theory and
experiment (including suitable corrections) is crude but
satisfactory.  It is important at this point (and entirely
feasible) to carry out similar analyses in other parts of the
phase diagram.  When more precision becomes available in future
experiments, more consistency checks will be possible (e.g., by
looking at $\left<a_2\right>$ and $\left<a_6\right>$), and it will be worthwhile to
include in the analysis the corrections of order
$k_{B}T/\omega$, which we have ignored herein.  It is clear
that gravitational corrections will have to be included
and that non-Gaussian effects will be important near instabilities.

\acknowledgments
We thank M. Jari\'{c}, J. K\"{a}s, L. Miao,
K. Ritchie, E. Sackmann, and T. Yeung for helpful discussions, and
 R. Lipowsky for generous support. Technical
 advice and help from A. Leung and W. Rawicz is appreciated.
 This work was supported in part by the Natural Sciences and Engineering
Research Council of Canada (H.-G.D. and M.W.) and by the
Medical Research Council of Canada (E.E.).


\begin{figure}
\vspace{2cm}
\caption{
  ADE phase diagram for $\alpha =1.4$, prolate region.  First-order
   boundaries ($D$) are indicated by solid lines; second-order
   boundaries ($C$), by dashed lines; and, spinodals ($M$), by dotted
   lines.  Lowest-energy shapes are illustrated for each region. All
   symmetry axes are vertical as indicated for the prolate. Prolate
   shapes are locally stable between the upper spinodal line
   $M^{pro}_{(0,-)}$ and the lower spinodal line $M^{pro}_{(2,+)}$.  The lines
$D^{pro/pear}$, $C^{pro/pear}$,
   $D^{pro/obl}$, and $D^{pro/nas}$ bound the region where prolates are
  the lowest-energy shapes.  In the region immediately above
  $D^{pro/pear}$ lowest-energy shapes are pear-like.  $L^{pear}$, is
  the limiting line at which the neck of the pear shape shrinks to
  zero radius producing a vesiculated shape, as indicated.  The
  lowest-energy states above $L^{pear}$ are dominated by vesiculated
  shapes.  In the region immediately below the prolates, oblate and
  non-axisymmetric (nas) shapes have lowest energy {\protect\cite{Jari95}}. The
  point T on the prolate/pear phase boundary is a tricritical point,
  separating first-order and second-order behavior.  CEP labels a
  critical end point, where a second-order boundary $C^{nas/obl}$ (not
  shown, since it is very close to $D^{pro/nas}$) disappears
  underneath the lower prolate boundary. Note for future reference
  that the limiting line $L^{pear}$ crosses the upper spinodal line
  $M^{pro}_{(0,-)}$ of the prolate phase at $v=0.875$ for  $\alpha =1.4$.
\label{fig:phasedia}}

\vspace{2cm}
\caption{
  The shape coefficient $a^{(0)}_4$ as a function of reduced volume
  $v$ for various values of the effective reduced spontaneous
  curvature $\bar c_0$, as indicated on the curves.  These curves
  allow us to infer a value of $\bar c_0$, if $a^{(0)}_4$ is known at
  given $v$.  This is the basis of the mapping procedure discussed in
  Sec. \ref{mapping}.  All the curves pass through $a^{(0)}_4=0$ at
  the sphere, $v=1$.  Although $a^{(0)}_4$ is almost independent of
  $\bar c_0$ at $v=0.85$, there is no common crossing point.
\label{fig:fish}}

\vspace{2cm}
\caption{
  Time-sequence of phase-contrast video images of vesicle A at
  $v=0.954$.  The images are ordered in time from the upper left to
  the lower right.  The elapsed time between images is 6.3 s.  The
  length of the long vesicle axis is approximately 20 $\mu$m.  The
  vesicle fluctuates about an axisymmetric prolate shape; however,
  each particular contour is different and, in general,
  non-axisymmetric.
\label{fig:seq95}}

\vspace{2cm}
\caption{
  Time-sequence of phase-contrast video images of vesicle A at
  $v=0.912$.  Times and scale are as in Fig.~\ref{fig:seq95}.  The
  vesicle is now more elongated than it was in Fig.~\ref{fig:seq95}.
  Strong pear-like fluctuations in each direction are now clearly
  visible.
\label{fig:seq91}}

\vspace{2cm}
\caption{
  Time-sequence of phase-contrast video images of vesicle A ($R_A =
  9.2\ \mu m$) at $v\simeq 0.878$, illustrating the budding process.
  The scale is as in Fig.~\ref{fig:seq95} and~\ref{fig:seq91}.  The
  time elapsed between images here is 1.2 s.  A pear-like fluctuation,
  much like those visible in Fig.~\ref{fig:seq91}, now carries the
  vesicle over the metastable barrier to the budded state.  The pear
  shaped contours correspond to {\em transient} shapes and are {\em
    not} stable. Note that the ratio of the vesicle size to the
  satellite size after budding is roughly 2.8.
\label{fig:budding}}

\vspace{2cm}
\caption{
  Time-sequence of phase-contrast video images of vesicle B ($R_A =
  5.5\ \mu m$) at $v\simeq 0.945$, illustrating the budding process
  for a smaller size vesicle.  The time elapsed between images is the
  same as in Fig.~\ref{fig:budding}. Here, the transition from the
  prolate via the transient pear to the budded state happens much
  more quickly than for vesicle A, due to the smaller hydrodynamic radius.
\label{fig:budding2}}

\vspace{2cm}
\caption{
  Typical time-series for the amplitudes $a_3$ and $a_4$ and for the
  effective reduced volume $v_e$ for vesicle A at reduced volume
  $\left<v_e\right> = 0.954$.  The dashed line corresponds to the mean
  amplitude $\left<a_n\right>$, which is close to zero for $a_3$
   {\protect\cite{comment15}}.  According to Eq. ({\protect\ref{equ:ident}}),
  the mean values $\left<v_{e}\right>$ and $\left<a_{4}\right>$ correspond to
  the zero-temperature quantities $v$ and $a^{(0)}_4$, respectively,
  which are the basis for the mapping.
\label{fig:tseries1}}

\vspace{2cm}
\caption{
  Same data as in Fig.~{\protect\ref{fig:tseries1}} but for reduced volume
  $\left<v_e\right> = 0.912$.  Note the longer time-scales for
  fluctuations in the $a_3$-mode as the shape instability (spinodal)
  is approached.
\label{fig:tseries2}}

\vspace{2cm}
\caption{
  Experimental trajectories without (Sec.~V, open symbols) and with
 (Sec.~VI, filled symbols)
  gravitational corrections in the $(v,\bar c_0)$ diagram
  for vesicles A, B, and C.  Corrected and uncorrected
  data points are joined by a thin line. As explained in
  the text, the vesicle follows a path from lower right to upper left,
  as it is heated. Vesicles A and B underwent a budding
  instability after the upper-left-most point of the trajectory,
  vesicle C budded from the point shown.  The spinodal lines
  $M^{pro}$ are the same as those shown in Fig.~{\protect\ref{fig:phasedia}}
  and
  mark the upper and lower boundaries of the region predicted by
 theory to be locally stable for prolate shapes.  Thus, for full
  consistency, the trajectories must terminate below the upper
  spinodal.  Vesicles B and C satisfy this criterion.  The
  final raw-data point on the trajectory of vesicle A is
  inconsistent with stability.  Including gravitational effects
  moves the thermal trajectories to smaller reduced volume.
  For  $g = 2.2$ (appropriate only for vesicle A),
 the instability boundaries are
  shifted by
  gravity, as shown.  The final point on the vesicle $A$
  trajectory becomes consistent with theory, only when, in addition to
  gravitational effects on the mapping, the gravitational tipping of
  the fluctuating pear-like shapes is incorporated by excluding a
  cluster of data points (square symbols), as discussed at the end of
  Sec.~{\protect\ref{results}}. Uncertainties are generally large near the
  spinodal lines and close to the sphere, where fluctuations become
  important.
 \label{fig:cmap}}

\vspace{2cm}
\caption{
  Experimental trajectories without (Sec. V, open symbols) and with
  (Sec. VI, filled symbols) gravitational corrections in the $(v,\bar
  m_0)$ phase diagram for vesicles A, B, and C using $\alpha=1.4$
  (see Fig.~\ref{fig:cmap} for a legend). The
  stability of data points does not depend on $\alpha$ and mirrors
  Fig.~\ref{fig:cmap}.  As for
  Fig.~\ref{fig:cmap}, the last point of the vesicle A trajectory
  becomes consistent with the theoretically calculated stability when
  corrected for gravitational effects (including the cluster
  exclusion, as discussed in the text).
  A theoretical thermal trajectory (with the simple form $\bar m_0 v = {\rm
  const}$) is shown for comparison.
  Although Figs.~\ref{fig:cmap}
  and \ref{fig:mmap} look similar, they are {\em not} connected
  by a simple rescaling of the vertical axis.
\label{fig:mmap}}

\vspace{2cm}
\caption{
  Computed gravitational corrections to the effective reduced volume
  $v_e$ at $g=2.2$, as is appropriate for vesicle A.
The shift ($v_e - v$) is shown
vertically as a function of
  $\bar c_0$ and $v$. The dashed line corresponds to the
  prolate-oblate transition. All shifts meet this line with a slope
  different from $-1$, as explained in the text. The shaded region
  ($v_e > 1$) is unphysical.
\label{fig:veff}}
\end{figure}

\begin{table}
\centering
\begin{tabular}{|c|lcc|} \hline
  Vesicle   &$T/\ $C$^\circ$& $\left<v_e\right>$  & $\left<a_4\right>$ \\
\hline
           &$28.7 \pm 0.01$       & $0.954 \pm 0.001$ &   $-0.0037 \pm 0.0011$
\\ \cline{2-4}
           &$32.7 \pm 0.01$       & $0.932 \pm 0.00$1 &    $0.0054 \pm 0.0013$
\\ \cline{2-4}
    A      &$37.8 \pm 0.01$       & $0.912 \pm 0.001$ &    $0.0038 \pm 0.0022$
\\ \cline{2-4}
           &$42.4 \pm 0.01$       & $0.894 \pm 0.001$ &    $0.0092 \pm 0.00$1
 \\ \cline{2-4}
           &$45.8 \pm 0.01\ $   budding      & $0.878 \pm 0.002$ &
          N/A          \\ \hline
           &$32.9 \pm 0.01$       & $0.970 \pm 0.001$ &    $0.005 \pm 0.001$
                \\ \cline{2-4}
    B      &$40.3 \pm 0.01$       & $0.950 \pm 0.001$ &    $0.016 \pm 0.001$
                   \\ \cline{2-4}
           &$42.0 \pm 0.01\ $   budding       &   $0.945 \pm 0.002$
               &           N/A          \\ \hline
    C      &$27.0 \pm 0.01\ $   budding    & $0.983 \pm 0.001$ &    $0.008 \pm
0.002$   \\ \hline
\end{tabular}
\vspace{2cm}
\caption{
Experimental results for vesicles $A$, $B$, and $C$. At each temperature
$T$, the values of $\left<v_e\right>$, and $\left<a_4\right>$ are shown.
The final budding temperature is also given, along with the 
(for A  and B extrapolated) reduced volume at budding.
\label{table}}
\end{table}


\begin{thebibliography}{99}
\bibitem{comment1}
The best previous example is probably Ref. \cite{Bern90}, where several
thermal trajectories were ``fit'' to the bilayer-couple model.  This
involved comparing observed and calculated shapes under the assumption of
constant integrated mean curvature $m$.  The parameters of
each trajectory were chosen to fit experimental shapes without any
independent confirmation.  Single video frames, chosen from what
was certainly a fluctuating ensemble, were taken to represent the
experimental shapes.
\bibitem{comment3}
The stretching modulus $K$ of a phospholipid membrane
is roughly 0.1 J/m$^{2}$, so
the energy necessary for significant change in area of a
micron-scale object is about $10^{-13}$J.  Similarly, the
energy necessary to resist significant osmotic pressure
is large on the scale of $\kappa$, so the volume of the vesicle
is effectively determined by osmotic balance between the inside
and the outside solutions.
\bibitem{comment4}
We regard $\kappa$, $\alpha$, $C_0$, and $D$ as material
parameters, unaffected by any shape changes.
\bibitem{comment5}
That is, except for the trivial Euclidean modes of translation
and rotation, any change in the shape $S$ which keeps $A$ and $V$
constant must cost energy.
\bibitem{comment7}
In the spontaneous curvature model, there is a prolate shape for any $\bar
c_0$.
 As $\bar c_0 \rightarrow \pm \infty$, these shapes reach limiting shapes with
finite
$m$ \cite{Bern90,Seif91}. The mapping equation (\ref{effectivecurvature}) then
implies
that in the ADE model a stationary prolate exists for any $\bar m_0$.
\bibitem{anomaly}
  Figure ~\ref{fig:fish} shows that $a^{(0)}_4(v=0.85,\bar c_{0})\simeq -0.021$
for all values of $\bar c_0$.
We have checked numerically and find that the common crossing
point is only approximate.  We do not understand the reason for
this near degeneracy.  What is clear, however, is that in the
region near $v=0.85$ measurement of $a^{(0)}_4$ for a prolate is not a good
predictor of $\bar c_0$.  Our data is at higher reduced
volumes, so this is not a problem.
\bibitem{comment8}
Structures with length scales below
0.2 $\mu$m are visible only
as interference patterns or not at all.  Nevertheless, we infer that such
structures are present, since we can
frequently see
their effects -- attachments, adhesions, etc. -- on the visible structures.
\bibitem{comment9}
This interval of 0.5 s is too coarse to follow dynamical
motion, except for the very slow modes which occur close to the
budding instability \cite{Doeb95}.  We could, of course,
retrieve the taped video images and achieve a resolution of 30
Hz, but not in real time.  However, for present purposes, what
we need is only a statistical sample of the full thermal shape
ensemble, so following the dynamics is unnecessary. We note that
our current set up allows now shape analysis at video frequency.
\bibitem{comment10}
Every so often the search algorithm loses the target vesicle between
successive frames, e.g., when when another vesicle passes
across the field.  Such events are identified by monitoring a
parameter like the overall contour length.  The difficult frame
is then discarded and the next available frame is grabbed for
analysis.
\bibitem{filter}
The coordinates of the contour were averaged 10 times
using  $ x^{n+1}_i = \frac{1}{4} (x^n_{i-1} + 2 x^n_i + x^n_{i+1})$,
and an analogous expression for $y$.
\bibitem{optics}
Note that we do not claim to be able to resolve structures which are only
30nm apart. It is only the {\em relative} position of the membrane which can be detected
with high precision.
\bibitem{comment12}
Where the prolate is the ground state, the ensemble is truly
stationary.  In regions of prolate metastability, there is an
effectively stationary restricted ensemble, as long as the
barrier height is appreciably larger than $k_{B}T$, as discussed
in Sec. \ref{phasediagram}.
\bibitem{comment13}
In principle, we can now calculate the corrections of order
$k_{B}T/\omega$.  In fact, we have not done so, because, in
situations where these corrections are appreciable, there are
other effects (non-Gaussian corrections, gravity) which are
also important and harder to compute.  For an improved level of
comparison between theory and experiment, it will be crucial to
take all these effects into account.
\bibitem{comment14}
This reversibility shows that
the neck remains open in the budded state.  The mechanism of
the energy barrier to reverse budding has not been clearly
identified.  It
may involve the intervention of short-range (e.g., van der
Waals) forces across the narrow neck.  ``De-necking'' seems always
to be accompanied by some
irreversible effects, i.e., one does not return precisely to the state of
the original vesicle at a lower $T$.
\bibitem{comment15}
Vesicle A exhibited a small but detectable up-down asymmetry of
unknown origin.  This manifested itself in a nonzero value of
$\left<a_3\right>$, even in the prolate phase.  For $v>0.92$ the observed value of
$\left<a_3\right>$ remained less than 0.005.  However, nearer the budding
instability, we observed $\left<a_3\right>\sim 0.04$, presumably because
the high ``susceptibility'' near the transition magnifies some
intrinsic asymmetry \cite{HGDthesis}.  None of these effects are visible in 
a casual scan of the video images.  One possible explanation is that vesicle A
incorporated some structural ``defect'' at
a scale below our resolution. Another possibility is that the gravitational
``tipping''
of the pear-fluctuations, which is discussed at the end of Sec.~\ref{results},
combined with too-short
a sampling time to average out rare events produces spurious non-zero odd
modes.
\bibitem{comment16}
Let the two sphere radii in suitable reduced units be $r_1$ and
$r_2$  with $r=r_{1}/r_{2}$.  The reduced area is
$1=r_{1}^{2}+r_{2}^{2}$ and the reduced volume is
$v=r_{1}^{3}+r_{2}^{3}$. Thus, we find $v = (1+r^3)/(1+r^2)^{3/2}$.
\bibitem{comment17}
These numbers reflect the relative densities of the interior and
exterior solutions, as prepared.  Small amounts of intermixing
may have occurred during the course of the experiments, which
would have the effect of decreasing g. We neglect the small variations in $g$
due to the temperature induced changes in $R_A$.
\bibitem{comment18}
Thermal fluctuations tend to make the average vesicle outline more circular,
so that we may expect that $\left<a_2\right>$ and $\left<a_4\right>$ go
to zero when fluctuations become sufficiently strong. This effect would 
lead to a larger inferred value of $a_4^{(0)}$, which (via Fig.~\ref{fig:fish})
would give a larger $\bar c_0$.
\end{thebibliography}
\end{document}